\documentstyle[preprint,eqsecnum,aps]{revtex}

\begin{document}
\bibliographystyle{prsty}
\draft

\title{Quantization of pseudoclassical model\\
        of spin one relativistic particle.}

\author{D. M. Gitman, A. E. Gon\c {c}alves}

\address{Instituto de F\'{\i}sica, Universidade de S\~ao Paulo\\
P.O. Box 20516, 01498-970 S\~ao Paulo, SP, Brazil}

\author{I. V. Tyutin}
\address{P.N. Lebedev Institute\\
53 Leninsky Pr., Moscow 117924, Russia}
\date{\today}

\maketitle

\begin{abstract}
A consistent procedure of canonical quantization of pseudoclassical model for
spin one relativistic particle is considered.
Two approaches to treat the quantization for the massless case are discussed,
the limit of the massive case and independent quantization of a modified
action. Quantum mechanics
constructed for the massive case proves to be equivalent to the Proca theory
and for massless case to the Maxwell theory.
Results obtained are compared with ones for the case of spinning
(spin one half) particle.

\end{abstract}

\section{Introduction}\label{an}

Classical and pseudoclassical models of relativistic particles and their
quantization are discussed lately in different contexts. One of the reason
is on these simple examples to learn how to solve some problems
which arise also in string theory, gravity and so on. On the
other hand it is interesting itself to find out whether there exist classical
models for any relativistic particles (with any spin), whose quantization
reproduces, in a sense, the corresponding field theory or one particle sector
of the corresponding quantum field theory.

A classical action of a scalar relativistic particle one can find, for
example, in the Landau text book \cite{Landau}. An action of spin one half
relativistic particle, with spinning degrees of freedom, describing by
anticommuting (grassmannian or odd) variables, was first proposed by
Berezin and Marinov
\cite{BM} and just after that discussed and investigated in papers
\cite{Casa,BCL,BDZVH,BVH,BSSW}.
Generalization of this model for particles with arbitrary spin was proposed
in \cite{Sri,GT}. The actions of the models obey different
kinds of gauge symmetry, in particular, of reparametrization invariance and
special supertransformations. Due to the reparametrizations in all the cases
Hamiltonian equal zero on the constraint surface. In the papers
\cite{BL,GNR,HA}, devoted to the quantization of these models, they tried to
avoid these difficulties, using the so called Dirac method of quantization of
theories with first-class constraints \cite{Dirac}, in which one considers
the first-class constraints in the sense of restrictions on the state vectors.
Unfortunately, in general case,
this scheme of quantization creates many questions, e.g. with Hilbert space
construction, what is Schr\"{o}dinger equation and so on. A consistent, but
more complicated technically way is to work in the physical sector, namely,
first, on the classical level, one has to impose gauge conditions to all the
first class-constraints to reduce the theory to one with second-class
constraints only, and then quantize by means of the Dirac brackets (we will
call such a method as canonical quantization). First canonical the
quantization for a
relativistic spin one half particle was done in \cite{DI}. In this paper we
are going to use this approach to quantize a relativistic particle
spin one. We consider a pseudoclassical model of relativistic
spin one particle both massive and massless with an action, which is
conventional generalization of Berezin-Marinov action, mentioned above, with
a Chern-Simons term. We impose gauge conditions to all the first class
constraints, except to one first-class constraint, which is quadratic in
fermionic
variables. In virtue of the structure of this constraint it is difficult, and
probably impossible without a reduction of the number of degrees of freedom,
to impose a conjugated gauge condition, on the other hand, treating this
constraint in the sense of restrictions of quantum states does not create
problems with Hilbert space construction. Thus, we quantize the theory
quasicanonically by means of Dirac brackets with respect to all other
constraints and  gauge conditions.
We demonstrate that quantum mechanics constructed is equivalent
to one-particle sector of the quantum theory of Proca vector field.
The quantization of the massless case is considered in two ways, as the limit
from the massive case and independently starting from the massless
Lagrangians without the variable $\psi^{5}$. For convenience, a
comparison with spin one half case is given.

\section{Pseudolassical models of spinning particles.}\label{bn}

A generalization of the pseudoclassical action of spin one-half
relativistic particle to the case of arbitrary spin $N/2$ can be
written in the form
\begin{eqnarray}\label{b1}
S &=& \int_{0}^{1}\left[
-\frac{1}{2e}\left(\dot{x}^{\mu}-i{\psi}_{a}^{\mu}{\chi}_{a}\right)
^{2} \right.
-\frac{e}{2}m^2-im\psi_{a}^{5}\chi_a  \nonumber  \\
&~&+ \left. \frac{1}{2}f_{ab}
\left(i\left[\psi_{an},~\psi^{n}_b\right]_{-}
+\kappa_{ab}\right)-i\psi_{an}\dot{\psi}_a^n\right]
d\tau\;,
\end{eqnarray}
where $x^\mu ,~e$ and $f_{ab}$ are even and $\psi_a^n,\;\chi_a$ are odd
variables ($f_{ab}$ is antisymmetric), dependent on a parameter
$\tau \in [0,1]$, which plays a role of time in this theory,
$\mu = {\overline{0,3}};~
a,b = {\overline {1,N}}~;~ n = (\mu,5)=\overline{0,3},5;~
\eta_{\mu \nu}={\rm diag}(1 -1 -1 -1); ~\eta_{mn}={\rm diag}(1 -1-1 -1 -1).$
Spinning degrees of freedom are described by odd (grassmannian)
variables $\psi_a^\mu$ and $\psi_a^5$;  odd $\chi_a$ and even
$e$ play an auxiliary role to make the action
reparametrization  and super gauge-invariant as well as to make it
possible consider both cases $m\neq 0$ and $m = 0$ on the same foot.
The summand $\frac{1}{2}\kappa_{ab}\int_0^1f_{ab}d\tau$, with even
coefficients $\kappa_{ab}$ plays the role of a Chern-Simons term and
can be added only in case  $N=2$ without
breaking of the rotational gauge symmetry \cite{HA}. Thus,
$\kappa_{ab} =\kappa\epsilon_{ab}\delta_{N,2}$ with an even
constant $\kappa$ and two dimensional Levi-Civita symbol
$\epsilon_{ab}$.

The are three types of gauge transformations under which the action
(\ref{b1}) is invariant:
reparametrizations
\begin{equation}\label{b2}
\delta x^{\mu} =  {\dot{x}}^{\mu} \xi\;,
\; \delta e = \frac{d}{d\tau}\left(e \xi\right)\;,
\; \delta f_{ab} = \frac{d}{d\tau}\left(f_{ab}\xi\right)\;,
\;\; \delta \psi_{a}^{n} =  \dot{\psi}_{a}^{n}  \xi\;,
\;\; \delta{\chi}_{a} = \frac{d}{d\tau}\left(\chi_a\xi\right)\;,
\end{equation}
supertransformations
\begin{eqnarray}\label{b3}
&~&\delta x^{\mu}  =  i\psi_{a}^{\mu}\epsilon_{a}\;,~~
\delta e =i\chi_{a}\epsilon_{a}\;,~~
\delta f_{ab} = 0\;,~~
\delta \chi_{a} = \dot{\epsilon}_{a}-f_{ab}\epsilon_{b}\;,\nonumber\\
&~&\delta \psi^{\mu}_{a}  = \frac{1}{2e}\left(\dot{x}^{\mu}-i\psi^{\mu}_{b}
\chi_{b}\right)\epsilon_{a}\;,~~
\delta \psi_{a}^{5} = \frac{m}{2}\epsilon_{a}\;,
\end{eqnarray}
and $O(N)$ rotations
\begin{equation}\label{b4}
\delta x^{\mu} =  0,\;
\delta e = 0,\;\delta f_{ab} = \dot{t}_{ab}+t_{ac}f_{cb}-
t_{bc}f_{ca},\;
\delta \psi_{a}^{n}  =  t_{ab}\psi_{b}^{n},\, \delta \chi_{a} =
t_{ab}\chi_{b}\;,
\end{equation}
with even parameters $\xi (\tau),\;t_{ab}(\tau)
=-t_{ba}(\tau)$,  and odd parameters $\epsilon_{a}(\tau)$.

Equations of motion have the form

\begin{eqnarray}
\frac{\delta S}{\delta x_{\mu}} & = &
\frac{d}{d\tau}\left[\frac{1}{e}\left(\dot{x}^{\mu}-i\psi^{\mu}
_{a}\chi_{a}\right)\right] = 0~,\label{b5}\\
\frac{\delta S}{\delta e} ~& = &  \frac{1}{2e^2}\left(\dot{x}^{\mu}
-i\psi_{a}^{\mu}\chi_{a}\right)^2-\frac{m^2}{2}=0~,\label{b6} \\
\frac{\delta S}{\delta f_{ab}} & = & \frac{1}{2}\left(
i\left[\psi_{na},~\psi_{b}^{n}\right]_{-}+
\kappa_{ab}\right) = 0\;,~~N\geq 2\;,\label{b7}\\
\frac{\delta_{r} S}{\delta \chi_{a}} ~& = & \frac{i}{e}
\left(\dot{x}^{\mu}-i\psi_{b}^{\mu}\chi_{b}\right)\psi_{a}^{\mu} -
im\psi_a^5 = 0\; ,\label{b8}\\
\frac{\delta_{r} S}{\delta \psi_{a\mu}} & = & 2i\left(\dot{\psi}_{a}^{\mu}
-f_{ab}\psi_{b}^{\mu}\right)-\frac{i}{e}\chi_a\left(\dot{x}^{\mu}-
i\psi_b^{\mu}\chi_b\right) = 0\; ,\label{b9}\\
\frac{\delta_{r} S}{\delta \psi_{a5}} & = & 2i\left(\dot{\psi}_{a}^{5}-
f_{ab}\psi_{b}^{5}-\frac{m}{2}\chi_a\right) = 0\; .\label{b10}
\end{eqnarray}
Calculating the total angular momentum $M_{\mu\nu}$, corresponding to the
action (\ref{b1}), we get

\begin{eqnarray}\label{b11}
&~&M_{\mu \nu} = L_{\mu \nu}+ S_{\mu \nu}\; ,\\
&~&L_{\mu \nu} = x_{\mu}p_{\nu}-x_{\nu}p_{\mu}\; ,\;\;
S_{\mu \nu} =i(\psi_{a\mu}\psi_{a\nu}-\psi_{a\nu}\psi_{a\mu})\; .\nonumber
\end{eqnarray}

\noindent The spatial part of $S_{\mu \nu}$ forms a tree-dimensional
spin vector ${\bf s}=( s^k)$
\begin{eqnarray}\label{b12}
s^k = \frac{1}{2}\epsilon_{kjl}S_{lj} = \sum_a s_a^k\; ,~~~
s_a^k = i\epsilon_{kjl}\psi_{al}\psi_{aj}\; ,
\end{eqnarray}
where $\epsilon_{kjl}$ is tree-dimensional Levi-Civita symbol
and there is no summation over $a$ in the last formula (\ref{b12}).
To demonstrate that this vector really behaves like  a spin one
should introduce an interaction with an external electromagnetic
field ${\it A}_{\mu}^{\it ext}(x)$ into the model and consider the
non-relativistic approximation. Unfortunately, in general case
it is impossible to introduce such an interaction in the action (\ref{b1})
in the same manner as for the spin one half\cite{BCL,BVH,BSSW}.
Namely, if one adds the terms

\begin{equation}\label{b13}
-g\dot{x}^{\mu}A_{\mu}^{ ext}
+igeF_{\mu \nu}^{ext}\psi_{a}^{\mu}\psi_{a}^{\nu}
\end{equation}
with arbitrary external field ${\it A}_{\mu}^{ext},\;F_{\mu \nu}
^{ext} = \partial_{\mu}A_{\nu}^{ext}-\partial_{\nu}A_{\mu}^{ext}$,
to the integrand (Lagrangian) of the expression (\ref{b1}), this Lagrangian
becomes inadmissible, i.e. the corresponding lagrangian equations
become
inconsistent for $N > 1$. One can check this by straightforward
calculations. However, if the external field is constant $(F_{\mu \nu}
={\rm const})$, the terms (\ref{b13}) can be added to the Lagrangian;
the equations of motion remain consistent, but the super gauge
symmetry (\ref{b3}) of the action disappears, namely equations of
motion have now only one
solution for $\chi ,\;  \chi = 0$. So, let us introduce the interaction
with a constant magnetic field $F_{0i}^{ext}=0,\;F_{ij}^{ext}
=-\epsilon_{ijk}B^k$, where $B^k$ are components of a magnetic
field {\bf B}, that is enough for our purposes.
Besides, we restore the velocity of light $c$ in the equations
of motion by the prescription $m\rightarrow mc$, $g\rightarrow g/c $
and impose two gauge conditions

\begin{equation}\label{b14}
\tau = x^0/c=t\; ,~~~~~~f_{ab}=0\;,
\end{equation}

\noindent to fix the gauge freedom, which
corresponds to the reparametrizations and $O(N)$ rotations.
\noindent Thus, we get in the case of consideration

\begin{eqnarray}
&~&\frac{1}{e^2}\left(\frac{dx^{\mu}}{dt}\right)^2-m^2c^2-\frac{2ig}{c}
F_{\mu \nu}^{ext}\psi_a^{\mu}\psi_a^{\nu} = 0\; ,\label{b15}\\
&~&\frac{d}{dt}\left(\frac{1}{e}\frac{dx_{\mu}}{dt}\right) =
\frac{g}{c}F_{\mu \nu}^{ext}\frac{dx^{\nu}}{dt}\; ,\;\;\;\dot{\psi}_{a\mu}=
\frac{eg}{c}F_{\mu \nu}^{ext}\psi_a^{\nu}\; ,\label{b17}\\
&~&\frac{1}{e}\left(\frac{dx_{\mu}}{dt}\right)\psi_a^{\mu}-
mc\psi_a^{5}=0\; ,\;\; \dot{\psi}_a^5=0\; ,\;\;\chi_a = 0\; ,\label{b19}\\
&~&\psi_{an}\psi_b^n = \frac{i}{2}\kappa_{ab}\; ,~~
N\geq 2\;.\label{b20}
\end{eqnarray}
In the limit $c \rightarrow \infty$ (${\bf B}/c$ - fixed) it follows from
the equation (\ref{b15}) that $e$ can be every-where replaced by
$1/m$. Then we obtain from (\ref{b17})
\begin{equation}\label{b21}
m\frac{d^2{\bf x}}{dt^2}=\frac{g}{c}\left[\frac{d{\bf x}}{dt}\times
{\bf B}\right]\; ,\;\;
\frac{d{\bf s}}{dt}=\frac{g}{mc}\left[{\bf s}\times{\bf B}
\right].
\end{equation}

\noindent It follows from the
equations (\ref{b19}) that $\psi_a^0 = \psi_a^5 = {\rm const}\;
,$ and therefore the constraint (\ref{b20}) takes the form
$\psi_a^i\psi_b^i =-i\kappa_{ab}/2\; .$ Using this,  one can calculate
$s_a^ks_b^k = 2(\psi_a^i\psi_b^i)^2 = \kappa^2
(1-\delta_{ab})\delta_{N,2}/2\;,$ so that
\begin{equation}\label{b23}
{\bf s}^2 = \left(\sum_as_a^k\right)^2 = \kappa^2\delta_{N,2}\;.
\end{equation}

\noindent Thus, one can interpret the equations (\ref{b21})
as describing the non-relativistic motion of a
charged particle with the total spin momentum {\bf s}, $({\bf s}^2
=\kappa^2\delta_{N,2})$, and with the total magnetic momentum
$g{\bf s}/mc$ in a constant magnetic field.

\section{Hamiltonian formulation. Constraints. Gauges
Conditions.}\label{cn}

Going over to the hamiltonian formulation, we introduce the
canonical momenta:

\begin{eqnarray} \label{c1}
&~&p_{\mu}~ =\frac{\partial L}{\partial \dot{x}_{\mu}} =
-\frac{1}{e}(\dot{x}_{\mu}-i\psi_{a\mu}\chi_{a})~,~~~
P_{e}~ = \frac{\partial L}{\partial \dot{e}} = 0\;,\nonumber\\
&~&P_{\chi_{a}} = \frac{\partial_{r}L}{\partial \dot{\chi}_{a}}
= 0\; ,~~~P_{an} = \frac{\partial_{r}L}{\partial \dot{\psi}^n_a} =
-i\psi_{an}\;,~~~
P_{f_{ab}} = \frac{\partial L}{\partial \dot{f}_{ab}} = 0\;.
\end{eqnarray}

\noindent It follows from (\ref{c1}) that there exist primary constraints
${\Phi}^{(1)} = 0$,
\begin{eqnarray}\Phi^{(1)} = \left\{
\begin{array}{lcl}
\Phi^{(1)}_{1a} &=& P_{\chi_{a}}\; ,\\
\Phi^{(1)}_{2} &=& P_{e}\; , \\
\Phi^{(1)}_{3an} &=& P_{an}+i\psi_{an}\; , \\
\Phi^{(1)}_{4ab} &=& P_{f_{ab}}\; .
\end{array}\right.\nonumber \end{eqnarray}
We construct the Hamiltonian $H^{(1)}$ according to the standard
procedure (we are using the notations of the book \cite{DI}),

\begin{eqnarray*}
H^{(1)} = H+\lambda_{A}\Phi_{A}^{(1)}\; ,~~ H =
\left(\frac{\partial_{r}L}{\partial\dot{q}}\dot{q}-L\right)
\left|_{\frac{\partial_{r}L}{\partial\dot{q}}=p}\right.\; ,~~
q = (x,e,\chi,\psi,f)\; ,
\end{eqnarray*}

\noindent and get for the $H$~:

\begin{equation}\label{c2}
H = -\frac{e}{2}(p^2-m^2)+i(p_{\mu}\psi_{a}^{\mu}+m\psi_{a}^{5})
\chi_a-\frac{1}{2}f_{ab}\left(i\left[\psi_{an},\; \psi^{n}_{b}
\right]_{-} + \kappa_{ab}\right).
\end{equation}

\noindent From the conditions of the conservation of the primary constraints
$\Phi^{(1)}$ in the time $\tau,\;\dot{\Phi}^{(1)}=\left\{ \Phi^{(1)},
H^{(1)}\right\}=0$, we find secondary constraints $\Phi^{(2)}=0$,

\begin{eqnarray}\label{c3}
\Phi^{(2)} = \left\{
\begin{array}{lcl}
\Phi^{(2)}_{1a} &=& p_{\mu}\psi_{a}^{\mu}+m\psi^{5}_{a}\; ,\\
\Phi^{(2)}_{2} &=& p^2-m^2\; , \\
\Phi^{(2)}_{3ab} &=& i\left[\psi_{an}\; ,\psi_{b}^{n}\right]_{-}
+\kappa_{ab}\; , \\
\end{array}\right. \end{eqnarray}

\noindent and determine $\lambda$, which correspond to
the primary constraint $\Phi^{(1)}_{3an}$. Thus, the
Hamiltonian $H$ appears to be proportional to the constraints, as one
could expect in the case of a reparametrization invariant theory,

\begin{equation}\label{c4}
H = -\frac{e}{2}\Phi^{(2)}_{2}+i\Phi^{(2)}_{1a}\chi_{a}-
\frac{1}{2}f_{ab}\Phi^{(2)}_{3ab}\; .
\end{equation}

\noindent No more secondary constraints arise from the Dirac procedure,
and the Lagrange's multipliers, corresponding to the primary
constraints $\Phi^{(1)}_{1a},\;\Phi^{(1)}_{2}$ and
$\Phi^{(1)}_{4ab}$, remain undetermined.
One can go over from the initial set of constraints $(\Phi^{(1)},
\Phi^{(2)})$ to the equivalent one $(\Phi^{(1)},{\widetilde{\Phi}}^{(2)})$,
where
\begin{eqnarray*}
\widetilde{\Phi}^{(2)} = \Phi^{(2)}\left|_{\psi_{an}\rightarrow
\widetilde{\psi}_{an} = \psi_{an}+\frac{i}{2}\Phi_{3an}^{(1)}\;.}
\right.
\end{eqnarray*}

\noindent The new set of constraints can be explicitly divided in a set of
the first-class constraints, which is $(\Phi^{(1)}_{1,2},
\Phi^{(1)}_{4ab}, \widetilde{\Phi}^{(2)})$, and in a set of second-
class constraints, which is $\Phi^{(1)}_{3an}$.
So, we are
dealing with a theory with first-class constraints. Our goal is
to quantize this theory. We choose the following way. We will
impose supplementary gauge conditions to all the first-class
constraints, excluding the constraint ${\widetilde{\Phi}}^{(2)}_{3ab}$.
As a result we will have only a set of first-class constraints,
which is reduction of ${\Phi}^{(2)}_{3ab}$ to the rest of constraints.
These  constraints
we suppose to use to specify the physical states according to Dirac
\cite{Dirac}. All other constraints will be of second-class and
will be used to form Dirac brackets.

Thus, let us impose preliminary the following gauge conditions:

\begin{eqnarray}\label{c5}
\left.
\begin{array}{ll}
\Phi^G_{1a} = \chi_{a} = 0\; , &
\Phi^G_{2ab} = f_{ab} = 0\; ,\\
\Phi^G_{3} = x_{0}-\zeta \tau = 0\; , &
\Phi^G_{4a} = \psi^{0}_{a} = 0\; ,
\end{array}\right.\end{eqnarray}

\noindent where $\zeta = -{\rm sign}~p_{0}$
(The gauge $x_0-\zeta\tau = 0$ was first proposed in papers
\cite{DI} as a conjugated gauge condition to the constraint
$p^2=m^2$ in the case of scalar and spinning particles. In
contrast with the gauge $x_0 = \tau$, which together with
the continuous reparametrization symmetry breaks the time
reflection symmetry and therefore fixes the variables $\zeta$,
the former gauge breaks only the continuous symmetry, so
that the variable $\zeta$ remains in the theory to describe
states of particles $\zeta = +1$ and states of antiparticles
$\zeta=-1$. Namely this circumstance allowed one to get
Klein-Gordon and Dirac equations as Schr\"{o}dinger ones in
course of the canonical quantization. To break the supergauge
symmetry the gauge condition
$\psi^5 = 0$  was used in \cite{DI}. In \cite{GG} the general
class of gauge conditions
of the form $\alpha\psi^0+\beta\psi^5 = 0$ was investigated
in case of $D-$dimensional spinning particles.)
The requirement of consistency of the constraint
$\Phi^G_{3},\;\dot{\Phi}^G_{3}=0$,
gives one more gauge condition

\begin{equation}\label{c6}
\Phi^G_{5} = e+\zeta p_{0}^{-1} = 0\; ,
\end{equation}
and the same requirements for the gauge condition (\ref{c5}),
(\ref{c6}) lead to the determination of the lagrangian multipliers, which
correspond to the primary constraints $\Phi^{(1)}_1$,
$\Phi^{(1)}_{2}$ and $\Phi^{(1)}_4$.

To go over to a time-independent set of constraints, we introduce
the variable ${x}_{0}',\;x_{0}' = x_{0}-\zeta\tau$,
instead of $x_{0}$ without changing the rest of the variables.
This is a canonical transformation in the space of all variables
with the generating function $W = x_{0}p_{0}'+\tau
\left|p_{0}'\right|+W_{0}$, where $W_{0}$ is the generating
function of the identity transformation with respect to all variables
except $x_{0},\;p_{0}$. The transformed Hamiltonian ${H^{(1)}}'$
is of the form
${H^{(1)}}' = H^{(1)} + \partial W/\partial \tau =
H + \left\{\Phi\right\}\; ,$
where $\left\{\Phi\right\}$ are terms proportional to
the constraints and $H$ is the physical Hamiltonian,

\begin{eqnarray}\label{c7}
H =\omega = \left({\bf p}^{2}+m^{2}\right)^{1/2}\; ,~~
{\bf p} = \left(p_k\right)\;.
\end{eqnarray}

We can present all the constraints of the theory (including the gauge
conditions), after the canonical transformation, in the following
equivalent form: $K = 0,\;\phi = 0,\;T = 0$,

\begin{eqnarray} \label{c8}
K &=& \left\{
\begin{array}{lllll}\chi_{a}\; ,& e - \omega^{-1}\; ,&x_{0}'\; ,
&f_{ab}\; , &\psi_a^0\; ,\\
P_{\chi_{a}}\; , &P_{e}\; ,&\left|p_{0}\right| - \omega\; ,
&P_{f_{ab}}\; ,&P_{a0}\;;
\end{array}\right. \\ \nonumber \\
\phi &=& \left\{
\begin{array}{ll}
p_i\psi_{a}^i+m\psi_{a}^{5}\; ,&\\
P_{al}+i\psi_{al}\; ,&l=1\;,2\;,3\;,5\;;\label{c9}
\end{array}\right. \\ \nonumber \\
T_{ab} &=& i\left(\left[\psi_a^i,~\psi_b^i\right]_{-}
+\left[\psi_a^5,~\psi_b^5\right]_{-}\right)
-\kappa_{ab}\; .
\label{c10}
\end{eqnarray}

\noindent The both sets of constraints $K$ and $\phi$ are of
second-class, only
$T$ are now first-class constraints. The set $K$ has the so
called special form \cite{DI}, in this case, if we eliminate the variables
$\chi,\;P_{\chi_{a}},\;e,\;P_{e},\;{x'}_{0}$,
$\left|p_{0}\right|,\;f_{ab},\;P_{f_{ab}}$ and $\psi_0^a$ from the
consideration, using these constraints, the Dirac brackets for the
rest of variables with respect to all the second-class constraints
$\left( K, \phi\right)$ reduce to ones with respect to the
constraints $\phi$ only.
Thus, we can only consider the variables $x^i,\;p_i$,
$\zeta,\;\psi_a^l,\;P_{al},\;l=(i,5)$  and two sets of
constraints, second-class one $\phi$ and first-class one $T$.
Often further we will use the transversal $\psi_a^{i\perp}$ and
the longitudinal $\psi_a^{{}^\parallel}$ parts of $\psi_a^i$, because of
\footnote{Here and further we are using the following notations
\begin{eqnarray*}
&~&a^{i\perp} = \Pi^i_j({\bf p})a^j\;,a^{i^\parallel}=
L^i_j({\bf p})a^j\;,
a^{{}^\parallel} = p_ja^j,\\
&~&\Pi^i_j({\bf p})+L^i_j({\bf p}) = \delta^i_j\;,
L^i_j({\bf p}) = p^{-2}p_ip_j \;,~~~p = \left|{\bf p}\right|\;.
\end{eqnarray*}}
these variables are convenient to treat both cases $m\neq 0$
and $m = 0$ on the same foot.
The first constraint (\ref{c9}) is, in fact,
a relation between $\psi_a^{{}^\parallel}$ and $\psi_a^5,\;
\psi_a^{{}^\parallel} = -m\psi_a^5\;,$
whereas $\psi_a^{i\perp}$ are not constrained.
Nonzero Dirac brackets between all the variables have the form

\begin{eqnarray}\label{c12}
&~&\left\{x^k,x^j\right\}_{D(\phi)} =
\frac{i}{\omega^2}\left[\psi_a^{k\perp},\psi_a^{j\perp}\right]_{-}
+\frac{im}{\omega^2p^2}\left(p_k\left[\psi_a^{j\perp},\psi_a^5
\right]_{-}-p_j\left[\psi_a^{k\perp},\psi_a^5\right]_{-}\right)\;,\\
&~&\left\{x^i,\psi_a^{j\perp}\right\}_{D(\phi)} =
-\frac{\psi_a^{i\perp}p_j}{p^2}+\frac{m}{p^2}\Pi^i_j\psi^5_a \;,
{}~~~\left\{x^i,\psi_a^5\right\}_{D(\phi)} =
-\frac{m}{\omega^2}\psi_a^{i\perp}
+\frac{m^2}{\omega^2p^2}p_i\psi_a^5\;,\nonumber\\
&~&\left\{\psi_a^{i\perp},\psi_a^{j\perp}\right\}_{D(\phi)} =
-\frac{i}{2}\delta_{ab}\Pi^i_j\;,~~~
\left\{\psi_a^5,\psi_b^5\right\}_{D(\phi)} =
-\frac{i}{2}\frac{p^2}{\omega^2}\delta_{ab}\;,\;\;
\left\{x^k, p_j\right\}_{D(\phi)} = \delta_j^k\;.\nonumber
\end{eqnarray}

To simplify the problem of quantization one can go over to
new variables, whose Dirac brackets are more simple. Namely,
let us introduce $\theta_a^i$~ and $X^k$, analogous to the
case of spin one half particles \cite{GG},
according to the formulas

\begin{eqnarray}\label{c13}
&~&X^k = x^k -\frac{i}{\omega+m}\left[\psi_a^{k\perp},\psi_a^5\right]_{-}
\;,~~~
\theta_a^i = \psi_a^{i\perp}-\frac{\omega}{p^2}p_i\psi_a^5\;;\nonumber\\
&~&x^i = X^i-\frac{i}{\omega\left(\omega+m\right)}\left[\theta_a^i,
\theta_a^{{}^\parallel}\right]_{-}~,~~~
\psi_a^{i\perp} = \theta_a^{i\perp}\;,~~~
\psi_a^5 = -\frac{1}{\omega}\theta_a^{{}^\parallel}\;.
\end{eqnarray}

\noindent Using the brackets (\ref{c12}), one gets

\begin{eqnarray}\label{c14}
&~&\left\{X^k,p_j\right\}_{D(\phi)} = \delta_j^k\;,~~~
\left\{X^k,X^j\right\}_{D(\phi)} = \left\{X^k,\theta_a^k\right\}
_{D(\phi)} = 0\;,\nonumber\\
&~&\left\{\theta_a^k,\theta_b^j\right\}_{D(\phi)} =
-\frac{i}{2}\delta_{ab}\delta_{kj}\; .
\end{eqnarray}

\noindent Variables $X^i,\;p_i,\;\zeta,\;\theta_a^k$, are
independent with respect to the second-class constraints (\ref{c9}).
Thus, on this stage we stay only with the first-class constraints
(\ref{c10}), which being written in the new variables $\theta_a^k$,
have the form

\begin{eqnarray}\label{c15}
T_{ab}  = i\left[\theta_a^k,\theta_b^k\right]_{-}-
\kappa_{ab}\;.
\end{eqnarray}

\noindent It is useful to adduce the expression for angular
momentum $M_{\mu\nu}$ in terms of the independent variables,

\begin{eqnarray}\label{c16}
M_{0j}&=&x_0p_j-x_jp_0=X_0p_j-X_jp_0-\frac{ip_o}
{\omega\left(\omega+m\right)}\left[\theta_a^j,\theta_a^{{}^\parallel}
\right]_{-}\;, \;\;  x_0 = \zeta\tau\;, ~~p_0=-\zeta\omega\;, \nonumber\\
M_{kj} &=& x_kp_j-x_jp_k+
i\left(\psi_a^{k\perp}\psi_a^{j\perp}-
\psi_a^{j\perp}\psi_a^{k\perp}\right)+\frac{2im}{p^2}
\left(p_k\psi_a^{j\perp}-p_j\psi_a^{k\perp}\right)\psi_a^5\nonumber\\
&=& X_kp_j-X_jp_k+i\left[\theta_a^k,\theta_a^j\right]_{-}\;,
\end{eqnarray}
\noindent One can check by straightforward calculations that
$M_{\mu\nu}$ together with $p_\mu$ form the Poincare algebra in sense
of Dirac brackets with respect to the constraints $\phi$,

\begin{eqnarray*}
\left\{M_{\mu\nu}, M_{\lambda\rho}\right\}_{D(\phi)}& =&
\eta_{\mu\lambda}M_{\nu\rho}-\eta_{\mu\rho}M_{\nu\lambda}+
\eta_{\nu\rho}M_{\mu\lambda}-\eta_{\nu\lambda}M_{\mu\rho}\; ,
\nonumber \\
\left\{p_{\mu}, M_{\nu\lambda}\right\}_{D(\phi)} &=&
-\eta_{\mu\nu}p_{\lambda}+\eta_{\mu\nu}p_{\nu}\; ,
{}~~\left\{p_{\mu},p_{\nu}\right\}_{D(\phi)}=0\; .
\end{eqnarray*}

\section{Quantization}\label{dn}

In the previous section we have imposed the gauge conditions to all the
first-class constraints except the set of constraint (\ref{c10}).
These constraints are quadratic in the fermionic variables.
On the one hand, that circumstance makes it difficult to impose a conjugated
gauge condition, on the other hand, imposing these constraints on  states
vectors does not creates problems with Hilbert space construction since the
corresponding operators of constraints have a discrete spectrum.
Thus, we suppose to treat only the constraints $T_{ab}$
in sense of the Dirac method. Namely, commutation relations between
the operators $\hat{X}^i,\;\hat{p}_i,\;\hat{\zeta},\;
\hat{\theta}_a^k$, which are related to the corresponding classical
variables, we calculate by means of Dirac brackets (\ref{c14}), so
that the  nonzero commutators are

\begin{eqnarray}\label{d1}
&~&\left[\hat{X}^k,\hat{p}_j\right]_{-} = i\left\{X^k,p_j
\right\}_{D(\phi)} = i\delta_j^k\;,\nonumber\\
&~&\left[\hat{\theta}_a^k,\hat{\theta}_b^j\right]_{+} =
i\left\{\theta_a^k,\theta_b^j\right\}_{D(\phi)} =
\frac{1}{2}\delta_{ab}\delta_{kj}\;.
\end{eqnarray}

\noindent We assume also the operator $\hat{\zeta}$ to have the eigenvalues
$\zeta = \pm 1$ by analogy with the classical theory, so that
\begin{eqnarray}\label{d2}
\hat{\zeta}^2 = 1\;.
\end{eqnarray}

Suppose ${\cal R}$ is a Hilbert space of vectors ${\bf f}
 \in {\cal R}$, where one can realize the relations (\ref{d1}),
(\ref{d2}). Then physical vectors have to obey the conditions
\begin{eqnarray}\label{d3}
i\left[\theta_a^k,\theta_b^k\right]_{-}{\bf f} =
\kappa_{ab}{\bf f}\;.
\end{eqnarray}
\noindent Besides, they have to obey the Schr\"{o}dinger equation
\begin{eqnarray}\label{d4}
\left(i\frac{\partial}{\partial\tau}-\hat{H}\right){\bf f}=0\;,
\end{eqnarray}
\noindent with the quantum Hamiltonian $\hat{H}$ constructed
according to the classical physical one (\ref{c7}),
\begin{eqnarray}\label{d5}
\hat{H}=\hat{\omega}=\left({\bf \hat{p}}^2+m^2\right)^{1/2}\;.
\end{eqnarray}
\noindent Going over to the physical time $x^0 = \zeta\tau$
(see \cite{DI})
one can transfer (\ref{d4}) to the form

\begin{eqnarray}\label{d6}
\left(i\frac{\partial}{\partial x_0}-\hat{\zeta}\hat{\omega}
\right){\bf f} = 0\;.
\end{eqnarray}

\noindent Hermitian operators of angular momentum $\hat{M}_{\mu\nu}$ can be
constructed according to the classical expression (\ref{c16}),

\begin{eqnarray}\label{d7}
\hat{M}_{0j}&=&\hat{X}_0\hat{p}_j-\frac{1}{2}\left[\hat{X}_j,
\hat{p}_0\right]_{+}
-\frac{i\hat{p}_o}{\hat{\omega}\left(\hat{\omega}+m\right)}
\left[\hat{\theta}_a^j,\hat{\theta}_a^{{}^\parallel}
\right]_{-}\;,\nonumber\\
\hat{M}_{kj} &=&
\hat{X}_k\hat{p}_j-\hat{X}_j\hat{p}_k+
i\left[\hat{\theta}_a^k,\hat{\theta}_a^j\right]_{-}\;.
\end{eqnarray}

\noindent In fact, all the formulas we adduced until this moment where written
for
arbitrary $N$. However, a realization of the relations (\ref{d1})
and (\ref{d2}) has to be considered separately for each $N$. In this
paper we suppose to emphasize the case of spin one, which corresponds
to $N=2$.  At the same time we believe that it is instructive to
compare this case with the case $N=1$, which can be quantized
completely canonically \cite{DI}. Thus, below we consider construction of
state spaces separately in two cases $N=1$ and $N=2$.

\subsection{Spin one half}

In this case $N=1$ and the first-class constraint $T_{ab}$ are absent.
We can construct the realization of the algebra (\ref{d1})
in the Hilbert space ${\cal R}$, whose elements ${\bf f}\in {\cal R}$
are four-component columns,

\begin{eqnarray}
{\bf f} =
\left(
\begin{array}{c}
f_1({\bf x})\\
f_2({\bf x})
\end{array}
\right),\nonumber
\end{eqnarray}

\noindent so that $f_1({\bf x})$ and $f_2({\bf x})$ are two
components columns.
We seek all the operators in the block-diagonal form, namely

\begin{equation}
\hat{\zeta}=\gamma^0\;, \;\;\;
\hat{p}_k = -i\partial_k {\bf I}\; ,~~~\hat{X}^k = X^k {\bf I}\;,~~~
\hat{\theta}^k = \frac{1}{2}\Sigma^k\; ,~~~~~~~~~~~~~~
\label{d8}
\end{equation}

\noindent where $\gamma^0$ is the zero gamma matrix, $I$ and ${\bf I}$
are $2\times 2$ and $4\times 4$ unit matrices, $\Sigma^k={\rm diag}(
\sigma^k, \sigma^k)$,  where $\sigma^k$ are Pauli matrices.
We interpret $f_{+}(x)=f_{1}(x)$ as the wave function of a
particle and $f_{-}(x)={\sigma}^2f^{\ast}_{2}(x)$ as that of an
antiparticle and define accordingly the scalar product in
${\cal R}$,
\begin{eqnarray}
\left({\bf f}, {\bf g}\right) = \int\left[f_1^{+}g_1+g_2^{+}f_2\right]
d{\bf x} = \int f^{+}_{\zeta}g_{\zeta}d{\bf x}\; , ~~\zeta=\pm~.
\label{d9}
\end{eqnarray}
The operators $\hat{X}^k,\;\hat{p}_k,\;\hat{\theta}^k,\;\hat{H}$ are
self-conjugate with respect to this scalar product. It follows from
(\ref{d6}) that

\begin{eqnarray}
i\frac{\partial}{\partial x_0}f_{\zeta} = \hat{\omega}f_{\zeta}.
\nonumber
\end{eqnarray}

\noindent Thus, in this case the equations for the wave functions
of a particle and antiparticle have the same form as it has to be in
the absence of an external electromagnetic field.

The operators of angular momentum (\ref{d7}) and the spin operator
$\hat{s}^k$ have the following  form in the realization in question

\begin{eqnarray}
&~&\hat{M}_{ij}=\hat{X}_i\hat{p}_j-\hat{X}_j\hat{p}_i-\frac{1}{2}
\epsilon_{ijk}\Sigma^k\; , \label{d10}\\
&~&\hat{M}_{0j}=\hat{X}_0\hat{p}_j-\hat{X}_j\hat{p}_0-\frac{i}{2}
\frac{\hat{p}_j}{\hat{p}_0}+\frac{\hat{p}_0}
{2\hat{w}(\hat{w}+m)}\epsilon_{jkl}\hat{p}_k\Sigma^l\; ,
\nonumber\\
&~&\hat{s}^k = i\epsilon_{kjl}\hat{\psi}^l\hat{\psi}^j =
\frac{1}{2}\Sigma^k.\nonumber
\end{eqnarray}

\noindent As it is known, the square of the Pauli-Lubanski vector
$\hat{W}^{\mu} = 1/2\epsilon^{\mu\nu\lambda\sigma}
\hat{M}_{\mu\nu}\hat{p}_{\sigma}$
 is a Casimir operator for the Poincare algebra.
For this realization and in the centre mass system

\begin{eqnarray*}
\hat{W}^0 = 0~,~~~\hat{W}^k =m\frac{\hat{p}_0}{\hat{\omega}}
\hat{s}^k\;,~~~
\hat{W}^2 = -\left(\hat{W}^i\right)^2 = -\frac{3}{4}m^2.
\end{eqnarray*}

\noindent The latter confirms that the system in question has
spin one half.

Now one can see that the quantum mechanics constructed is
completely equivalent to the standard Dirac theory, namely it
is connected with the latter by the unitary Foldy-Wouthuysen
transformation \cite{FW}. Doing this transformation in the equation
(\ref{d6}), we are coming to the standard Dirac equation (see \cite{DI}),
$$
{\bf  f} = {\cal U}\Psi\; ,~~~~
{\cal U}=\frac{\hat{\omega}+m+\mbox{\boldmath $\gamma $}{\bf\hat p}}
{(2\hat{\omega}(\hat{\omega}+m ))^{1/2}}\; ,
{}~~~~\left(i\gamma^{\mu}\partial_{\mu}-m\right)\Psi = 0 \;.
$$
Besides, applying the same transformation to the operators (\ref{d10}),
we get the operators of the angular momentum in the Dirac theory
\cite{GG},
\[
{\cal U}^{+}\hat{M}_{\mu\nu}{\cal U} = \hat{X}_{\mu}\hat{p}_{\nu}
-\hat{X}_{\nu}\hat{p}_{\mu}- \frac{1}{2}\sigma_{\mu\nu}\;,~~~
\sigma_{\mu\nu} = \frac{i}{2}\left[\gamma_{\mu},\gamma_{\nu}
\right]_{-}\;.
\]

\subsection{Spin one}

\noindent The relations (\ref{d1}) (\ref{d2}) for $\hat{X}^k,\;
\hat{p}_j$ and $\hat{\zeta}$ we can realize in a Hilbert space
${\cal R}_{scal}$, whose elements are two-component columns
$f\in {\cal R}_{scal}$,

\begin{eqnarray}
f =
\left(
\begin{array}{c}
f_1({\bf x})\\
f_2({\bf x})
\end{array}
\right)\; ,~~~~~~~~~~~f_{\zeta}({\bf x}) \in L_2\; ,~~\zeta = 1,2\; ,
\nonumber
\end{eqnarray}
in the following natural way \cite{DI}:

\[
\hat{\zeta}=\sigma^3\;, \;\;\;
\hat{X}^k = x^kI\; ,\;\;\;\hat{p}_k= -i\partial_kI\;.
\]

\noindent The scalar product in
${\cal R}_{scal}$ we select in the form

\begin{eqnarray}\label{d12}
\left(f, g\right) = \int\left[f_1^{\ast}g_1+g_2^{\ast}f_2\right]
d{\bf x}\; .
\end{eqnarray}

\noindent The commutation relations (\ref{d1}) for
$\theta_a^k,~ a=1,~2$, we realize in a Hilbert space ${\cal R}_{spin}$,
which is a Fock space constructed by
means of tree kinds of Fermi annihilation and creation operators
$b_k,\;b_k^{+},\;k=1,~2,~3$,

\begin{eqnarray}
&~&\theta^k_1 = \frac{1}{2}\left(b_k + b_k^{+}\right),\; ~~~
\theta^k_2 = \frac{i}{2}\left(b_k^{+} -b_k\right)\; ,\label{d13}\\
&~&\left[b_k,b_j^{+}\right]_{+} = \delta_{kj}\; ,~~~~~~~
\left[b_k,b_j\right]_{+} = \left[b_k^{+}\; ,b_j^{+}\right]_{+}=0\; .
\nonumber
\end{eqnarray}

\noindent Due to the Fermi statistics of these operators
the space ${\cal R}_{spin}$ is finite-dimensional space of
vectors $v \in {\cal R}_{spin}$, with basis
vectors $v^{(0)}\;,v^{(1)}\;,v^{(2)}\;,v^{(3)}$,

\begin{eqnarray}\label{d14}
&~&v^{(0)} = |0>\; ,~~~b_k^{+}|0> = 0\;, k=1,2,3\;,\nonumber\\
&~&v_k^{(1)} = b_k^{+}|0>~~~\;,
v_{k}^{(2)} =\frac{1}{2} \epsilon_{kji}b_j^{+}b_i^{+}|0>\;,~~~
v^{(3)} =\frac{1}{6}\epsilon_{ijk} b_i^{+}b_j^{+}b_k^{+}|0>\; ,
\end{eqnarray}

\noindent which are eigen for the operator
$\hat{n}=\sum_kb_k^{+}b_k$,

\begin{eqnarray}\label{d15}
\hat{n}v^{(n )} = nv^{(n)}\; ,~~~~n = 0\; ,1\; ,2\; ,3\; .
\end{eqnarray}

\noindent The total Hilbert space ${\cal R}$ is the direct product
of ${\cal R}_{scal}$ and ${\cal R}_{spin}$.

Calculating the operators of angular momentum $\hat{M}_{\mu\nu}$,
spin $\hat{s}^k$ and square of Pauli-Lubanski vector
in the realization, we get

\begin{eqnarray}\label{d15a}
&~&\hat{M}_{ij}=\hat{X}_i\hat{p}_j-\hat{X}_j\hat{p}_i +
i\left(b_ib_j^{+}-b_jb_i^{+}\right)\; ,\nonumber\\
&~&\hat{M}_{0j}=\hat{X}_0\hat{p}_j-\frac{1}{2}
\left[\hat{X}_j,\hat{p}_0\right]_{+} +
\frac{\hat{p}_0}{\hat{w}(\hat{w}+m)}
\left(\hat{p}_kb_kb_j^{+}-b_j\hat{p}_kb_k^{+}\right)\; ,\\
&~&\hat{s}^k = \frac{i}{2}\epsilon_{kjl}\left(b_j^{+}b_l-
b^{+}_lb_j\right)\;,~~~
\hat{W}^2 = -m^2\hat{n}\left(3-\hat{n}\right)\;.\label{d16}
\end{eqnarray}

\noindent The operator $\hat{n}$ commutes with $\hat{H}$,~
$\hat{p}_{\mu}$ and $\hat{M}_{\mu\nu}$, that means that states
with a fixes $n$ form invariant subspaces.
In this realization the equation (\ref{d3}) imposes only
restrictions on the vectors $v$ from ${\cal R}_{spin}$,

\begin{eqnarray*}
\hat{n}v = \left(\kappa+\frac{3}{2}\right)v\;,
\end{eqnarray*}

\noindent they have to be eigenstates of the operator $\hat{n}$.
That implies that $\kappa$ takes on the values $-3/2$,
$-1/2,\;1/2,\;3/2$.
Due to (\ref{d16}) theories with $\kappa = \pm 1/2$
describe particles
with spin one, whereas theories with $\kappa = \pm 3/2$
describe spinless particles. The canonical quantization of the
latter case was described in \cite{DI}, thus, we consider here only
the former case. First, let us take $\kappa = -1/2$. In this case
$n=1$ and a general form of the time dependent state vector ${\bf f}
\in {\cal R}$ is

\begin{eqnarray}\label{d17}
{\bf f} = v_k^{(1)}f^k(x)\;.
\end{eqnarray}

\noindent Due to (\ref{d6}) each component $f^k(x)$ obeys the Klein-
Gordon equation,

\begin{eqnarray}\label{d18}
\left(\Box +m^2\right)f^k(x)  = 0\;,~~\Box = \partial_{\mu}
\partial^{\mu}\;.
\end{eqnarray}

\noindent We interpret $f_{(+)}^k(x) = f_1^k(x)$ as the wave function
of a particle and $f_{(-)}^k(x) = f_2^{k\ast}(x)$
as the wave function of antiparticle with spin one.
According to (\ref{d12}) the scalar product of two
state vectors has the following form

\begin{eqnarray}\label{d19}
\left({\bf f}, {\bf g}\right) = \int\left[f_1^{k\ast}g_1^k+
g_2^{k\ast}f_2^k\right]
d{\bf x} = \int f^{k\ast}_{(\zeta)}g_{(\zeta)}^kd{\bf x}\; ,
{}~~\zeta=\pm~.
\end{eqnarray}

Now one can find a correspondence between the quantum
mechanics constructed and the classical Proca field,
which describe particles
of spin one in the field theory. To this end we construct a vector
field ${\cal A}_{\mu}(x)$ from the functions $f^k(x)$ in the following way

\begin{eqnarray}\label{d20}
{\cal A}_{\mu}(x) = \frac{1}{\sqrt{2\hat{\omega}}}\xi_{\mu}^{(k)}
(\hat{{\bf p}})
\left(f_1^k(x)+f_2^k(x)\right)\;,
\end{eqnarray}

\noindent with polarization vectors $\xi_{\mu}^{(k)}({\bf p})$,  having
the form

\begin{eqnarray}\label{d21}
&~&\xi_0^{(k)}({\bf p}) = \frac{p_0p_k}{m\omega}\;,~~
\xi_i^{(k)}({\bf p}) = \delta_i^k + \frac{p_ip_k}{m\left(m+\omega\right)}\;,
\\
&~&\xi_{\mu}^{(k)}({\bf p})p^{\mu} = 0\;,~~
\xi_{\mu}^{(k)}({\bf p})\xi_{\nu}^{(k)}({\bf p}) = -\eta_{\mu\nu}+
\frac{p_{\mu}p_{\nu}}{m^2}\;,~~p_0 = -\zeta\omega\;.
\label{d22}
\end{eqnarray}

\noindent One can check, using (\ref{d18}) and
(\ref{d20}), that the field ${\cal A}_{\mu}(x)$ obeys the equations

\begin{equation}\label{d23}
\left(\Box +m^2\right){\cal A}_{\mu}(x)  = 0\;,~~\partial_{\mu}
{\cal A}^{\mu}(x) = 0\;,
\end{equation}

\noindent which are just equations for the Proca field \cite{BS}. Moreover,
one can find the action of the generators (\ref{d7}) on the field
${\cal A}_{\mu}(x)$, calculating their action on the vector
(\ref{d17}) and using the representation (\ref{d20}),

\begin{eqnarray}\label{d24}
\hat{M}_{\alpha\beta}{\cal A}_{\mu}(x) =
\left(x_{\alpha}p_{\beta}-x_{\beta}p_{\alpha}\right){\cal A}_{\mu}(x)-
i\left(\eta_{\alpha\mu}{\cal A}_{\beta}(x)-\eta_{\beta\mu}
{\cal A}_{\alpha}(x)
\right)\;,~~p_{\alpha}  = -i\partial_{\alpha}\;.
\end{eqnarray}

\noindent That result reproduces the transformation properties
of a vector field under the Lorentz rotations with $\delta x^{\mu}=
\omega^{\mu\nu}x_{\nu}$,

\begin{eqnarray}\label{d25}
\delta {\cal A}_{\mu}(x) = \frac{i}{2}\hat{M}_{\alpha\beta}{\cal A}_{\mu}
(x)\omega^{\alpha\beta}\;.
\end{eqnarray}

\noindent It is also instructive to point out a correspondence
between the quantum mechanics constructed and one particle sector
of the quantum theory of the Proca field. In this  quantum theory the
Proca field appears to be the operator

\begin{eqnarray*}
\hat{A}_{\mu}(x) = \int\frac{d{\bf p}}{\sqrt{2\omega\left(2\pi
\right)^3}}\left[e^{-ipx}a_k\left({\bf p}\right)\xi_{\mu}^{(k)}
\left({\bf p}\right) +
e^{ipx}d_k^{+}\left({\bf p}\right)\xi
_{\mu}^{(k)\ast}\left({\bf p}\right)\right]\;,
\end{eqnarray*}

\noindent where $a_k({\bf p})$,~$a_k^{+}({\bf p})$,~$d_k({\bf p})$,
{}~$d_k^{+}({\bf  p})$,~$k = 1,~2,~3$~ are two kinds of Bose,
annihilation and creation operators,
$p_0 = \omega$, and the polarization vectors $\xi_{\mu}^{(k)}
({\bf p})$ obey just the conditions (\ref{d22}). If we choose for them real
expressions (\ref{d21}), then the relations hold

\begin{eqnarray*}
&~&{\cal A}_{\mu}(x) =  <0|\hat{A}_{\mu}(x)|f_1>+<f_2|\hat{A}_{\mu}(x)|0>\;,\\
&~&|f_1> = \int d{\bf p}\tilde{f}_1^k({\bf p})a_k^{+}({\bf p})|0>\;,~~
\tilde{f}_1^k({\bf p}) = \left.\int \frac{d{\bf x}}{\left(2\pi\right)^
{(3/2)}}e^{-i{\bf px}}f_{(+)}^k(x)\right|_{x^0=0}\;,\\
&~&|f_2> = \int d{\bf p}\tilde{f}_2^k({\bf p})d_k^{+}({\bf p})|0>\;,~~
\tilde{f}_2^k({\bf p}) = \left.\int \frac{d{\bf x}}{\left(2\pi\right)^
{(3/2)}}e^{-i{\bf px}}f_{(-)}^k(x)\right|_{x^0=0}\;,
\end{eqnarray*}

\noindent so that ${\cal A}_\mu(x)$ is the classical Proca field (\ref{d20}).
In fact, by such a choice of the polarization vectors, we
have a direct correspondence between the wave functions of particles
and antiparticles $f^k_{(\zeta)}$
in the quantum mechanics and the states $|f_{1,2}>$
in the quantum field theory.

Finally, let us consider the case $\kappa = 1/2$, which also
describes a particle spin one. In this case $n=2$ and a general form
of the time dependent state vector ${\bf f} \in {\cal R}$ is

\begin{eqnarray}\label{d26}
{\bf f}  =v_k^{(2)} f^k(x)\;.
\end{eqnarray}

\noindent On can check by straightforward calculations that $f^k(x)$
from the eq. (\ref{d26}) obeys the same equations and appears in the same form
in all the constructions as $f^k(x)$ from the eq. (\ref{d17}). Moreover, the
action of the generators $\hat{M}_{\mu\nu}$ on the basis vectors
$v_i^{(1)}$ and $v_i^{(2)}$ is equal.
That provides equal transformation properties
for the field (\ref{d18}) constructed by $f^k(x)$ in both cases. All that
testifies that both theories with $\kappa = \pm 1/2$
describe spin one particles.

\section{Massless case. Quantum mechanics of photon. }\label{en}

Here we are going to discuss the problem of quantization of massless
particles spin one half and spin one. In this connection, one can consider the
limit $m=0$ of the above constructed quantum mechanics
and compare it with an independent quantization  of classical
action, describing massless particles at the beginning. As to the
limit, one can remark that all formulas are nonsingular in the mass and admit
such a limit. On the classical level, after the gauge fixing,
it is possible to use, on the surface of the second-class constraints,
 the variables $x^i,\;p_i,\;\zeta,
\;\psi_a^{i\perp},\;\psi_a^5$ or the variables $X^i,\;p_i,
\;\zeta,\;\theta_a^i$, the
Dirac brackets of the latter do not contain mass at all and expressions of
the former via the latter are nonsingular in the mass. The first set of the
variables at $m=0$  splits into two (anti)
commuting one with another groups $x^i,\;p_i,\;\psi_a^{i\perp}$,
and $\psi_a^5$. The Poincare generators are only
expressed via the first group of variables and commute with
$\psi_a^5$. Instead of the Casimir operator $W^2$, which vanishes at $m=0$,
appears a new one, helicity $\Lambda$,

\begin{eqnarray}\label{e1}
\Lambda = \hat{p}^{-1}\hat{p}_k\hat{s}^k\;,
\end{eqnarray}

\noindent It turns out that at $m = 0$ the variable $\psi_a^5$
can be omitted from the action (\ref{b1}). The quantization of such modified
action reproduces the physical sector (in particular, quantum
mechanics of the transversal photons)
of the limit of the massive quantum
mechanics. Below we adduce  details of the limit $m=0$ for two cases: of
spin one half and spin one, taking into account general properties mentioned
above, and emphasizing mainly differences from the massive case.

\subsection{Massless particle spin one half}

As we have mentioned above, the Dirac brackets for the variables
$X^i,\;p_i,\;\zeta,\;\theta_a^k$ do not depend on the mass,
that means that realization (\ref{d1}), (\ref{d2}) remains in the limit
$m=0$. It is clear that the realization does not depend on the
presence of the operator $\hat{\psi}^5$.
In the limit we have $\hat{\psi}^5 =i \hat{p}^{-1}\hat{p}_k \epsilon_{kjl}
\hat{\psi}^{l\perp}\hat{\psi}^{j\perp} = \Lambda\;,$ where $\Lambda$ is the
helicity operator. The Schr\"{o}dinger equation (\ref{d6}) with $m = 0$
gives the Dirac equation
with $m = 0$ after the corresponding FW transformation. The total
Hilbert space forms now a reducible representation of the Poincare group
(right and left neutrinos). It follows from the described structure
of the quantum mechanics that in the limit $m=0$
one does not need the variable $\psi^5$ at the theory. Indeed,
one can take the action (\ref{b1}) at $m=0$ and omit
$\psi^5$ in the beginning. In such a theory, after the same gauge
fixing (in particular, $\psi_0=0$), we
have only the variables $x^i,\;p_i,\;\zeta, \;\psi^{i\perp}$ on the
constraint surface. Their Dirac brackets and the expressions of the
Poincare generators coincide with the corresponding expressions
of the massive theory at $m=0$. The same realization is
available. If one introduces the operator
$i\hat{p}^{-1}\hat{p}_k\epsilon_{kjl}\hat{\psi}^{l\perp}
\hat{\psi}^{j\perp}$, which is in fact the operator $\hat{\psi}^5$ of the
massive case, then the theory literally coincides with the limit
of the massive case. In this connection one can remark that the
dimensionality of the Hilbert space in the discussed realization
does not depend on the presence of the variable $\psi^5$ at
$m=0$ and coincide with dimensionality of the massive case.

\subsection{Quantum mechanics of photon}

Now let us turn to the massless case $N=2$, which,
according to our expectations
has to describe a photon. First, we consider the limit $m=0$ of the
massive spin one case with $\kappa=-1/2$. According to our interpretation,
states with $\zeta=+1$ correspond to particles and with $\kappa=-1$ to
antiparticles. Because of our aim is a photon, which is neutral, we may
restrict ourselves to consider the limit of massive quantum
mechanics of neutral spin one particle. To get such a quantum mechanics one
needs to replace the gauge condition $x_{0}=\zeta\tau$ by the
one $x_{0}=\tau$, the latter fixes, besides the reparametrization
gauge freedom,
the discrete variable $\zeta \; (\zeta=1) $ as well \cite{DI}. Thus, the
operator $\hat{\zeta}$ disappears
from the consideration and elements of the ${\cal R}_{scal}$ are
merely functions $f({\bf x})$ from $L_2$ with the scalar product
$(f,g)=\int f^{\ast}gd{\bf x}$. The realization for
$\hat{X}^i=\hat{x}^i,\;\hat{p}_i,\;\hat{\theta}_a^k$ remains
the same as at $m\neq 0$. The operator of helicity and its square
have the form

\begin{eqnarray*}
\Lambda = i\hat{p}^{-1}\hat{p}_i\epsilon_{ijl}b_j^{+\perp}b_l^{\perp},\;\;
\Lambda^2 = \hat{n}^{\perp}\left(2-\hat{n}^{\perp}\right)\;,
\;\; \hat{n}^{\perp} = b_j^{+\perp}b_j^{\perp}\;.
\end{eqnarray*}

\noindent The total Hilbert space splits into the two
invariant subspaces, with
$\Lambda^2 = 1$, $\Lambda = \pm 1$ and with $\Lambda = 0$. The
first subspace can be created by the operators
$\hat{x}^i,\;\hat{p}_i,\;\hat{\psi}_a^{i\perp} =\hat{\theta}_a^{i\perp}$,
whereas the second one by the operators $\hat{x}^i,\;\hat{p}_i,
\;\hat{\psi}_a^5 =-\hat{p}^{-1}\hat{\theta}_a^{{}^\parallel}$. We
treat the subspace with $\Lambda^2 = 1$ as the Hilbert space of
transversal photons with helicity $\Lambda = \pm 1$. The subspace
with $\Lambda = 0$ we treat as the Hilbert space of longitudinal
photons with helicity $0$. To exclude the longitudinal photons from the
consideration one needs to impose a supplementary condition $\Lambda^2 = 1$.
On the other hand, to get a theory, containing only the transversal
photons, one can start from the action (\ref{b1}) $N=2$, $m=0$,
without the variables $\psi_a^5$,
\[
\int_{0}^{1}\left[
-\frac{1}{2e}\left(\dot{x}^{\mu}-i{\psi}_{a}^{\mu}{\chi}_{a}\right)
^{2} +\frac{i}{2}f_{ab}\left[\psi_{a\mu},~\psi^{\mu}_b\right]_{-}
-i\psi_{a\mu}\dot{\psi}_a^\mu\right]d\tau\;.
\]
In this case one can have the
same realization for the operators $\hat{x}^i,\;\hat{p}_i$,
$\hat{\psi}_a^{i\perp}$ as in quantum mechanics with $\psi_a^5$
at $m=0$. Instead of the operator $\hat{n}$ in the condition (\ref{d3})
appears the operator $\hat{n}^{\perp}$,

\begin{eqnarray}\label{e2}
\hat{n}^{\perp}{\bf f} = \left(\kappa+1\right){\bf f}\;.
\end{eqnarray}

\noindent Its eigenvalues $n^{\perp}$ can be only  $0,\;1,\;2$, so that
$\kappa$ takes now on the values $0$, $\pm 1$. The cases $n^{\perp}=0,~2;\;
\kappa = \pm 1$ correspond to the spinless particles; the case
$n^{\perp}=1;\;\kappa = 0$ corresponds to the limit $m=0$ of
the quantum theory with the action (\ref{b1}) with $\kappa =
-1/2$, sector $\Lambda^2 = 1$, and reproduces the quantum
mechanics of the transversal photons.

Finally, we can demonstrate that the quantum mechanics of the transversal
photons reproduces in a sense the classical Maxwell theory and is equivalent
to one-particle sector of quantum theory of Maxwell field. To this end
let us rewrite the representation (\ref{d17}) in the form

\begin{eqnarray*}
{\bf f}=  v^{(1)\perp}_kf^{k\perp}(x)+v^{(1)^\parallel}_k
f^{k^\parallel}(x)\;,
\end{eqnarray*}

\noindent where the transversal and longitudinal components are defined by
means of the corresponding projectors,
$\Pi_k^j(\hat{{\bf p}}),\; L_k^j(\hat{{\bf p}})$. After the limit $m=0$
one can interpret $f^{k\perp}(x)$  as the wave function of
transversal photons. To construct the classical electromagnetic field we
have to use the wave functions
$f^{k\perp}(x)$ in the same way we had used the wave functions
$f^{k\parallel}(x)$ in the previous section to construct
the Proca field. Namely, we define  a vector field ${\cal A}_\mu(x)$ in the
following way
\begin{eqnarray}\label{e3}
{\cal A}_\mu(x)  = \frac{1}{\sqrt{2\hat{\omega}}}\xi_\mu^{(k)\perp}
\left[f^{k\perp}(x)+f^{k\perp\ast}(x)\right]\;,
\end{eqnarray}

\noindent where $\xi_\mu^{(k)\perp}$ are transversal components of
the polarization vectors (\ref{d21}),

\begin{eqnarray}\label{e4}
\xi_\mu^{(k)\perp} = \delta^i_\mu\Pi^{k}_{i}(\hat{{\bf p}})\; .
\end{eqnarray}

\noindent Due to the equation (\ref{d18}) and the
structure of the polarization
vectors (\ref{e4}), the field (\ref{e3}) obeys the Maxwell equations in the
Coulomb gauge,

\begin{eqnarray*}
\Box^2{\cal A}_\mu(x) = 0\;,~~~\partial_j{\cal A}^j(x)=0\;,~~~
{\cal A}_0(x) = 0\;.
\end{eqnarray*}

Let us turn to the quantum theory of the Maxwell field. In the
Coulomb gauge the operator of the vector potential has the form

\begin{eqnarray*}
\hat{A}_k(x) = \int\frac{d{\bf p}}{\sqrt{2p_0\left(2\pi\right)^3}}
\left[e^{-ipx}c_{\lambda}({\bf p}) +e^{ipx}c^+_{\lambda}({\bf p})
\right]e_k^{(\lambda)}({\bf p})\;,~~~\lambda = 1,\;2\;,~~~
p_0 = \left|{\bf p}\right|\;,
\end{eqnarray*}

\noindent where $c_\lambda^+({\bf p})\;,~c_\lambda({\bf p})$
are  creation and annihilation operators of transversal
photons and $e_k^{(\lambda)}({\bf p)}$ are two polarization vectors,
which are selected here to be
real,

\begin{eqnarray*}
e_k^{(\lambda)}({\bf p)}e_k^{(\lambda')}({\bf p)} =
\delta_{\lambda\lambda'}\;,~~~e_k^{(\lambda)}p_k = 0\;.
\end{eqnarray*}

\noindent Classical vector potential ${\cal A}_k(x)$ can be
constructed as

\begin{equation}\label{e5}
{\cal A}_k(x)=<0|\hat{A}_k(x)|f> + <f|\hat{A}_k(x)|0>\;,
\end{equation}
\[
c_\lambda^{+}({\bf p})|0> = 0\;,~~
|f> = \int{\tilde f}^{(\lambda)}({\bf p)}c_\lambda^{+}({\bf p)}|0>
\;,~~{\tilde f}^{(\lambda)}({\bf p)} = \left.\int
\frac{d{\bf p}}{\left(2\pi\right)^{3/2}}e^{-i{\bf px}}e^{(\lambda)}_k
({\bf p})f^{k\perp}(x)\right|_{x^0= 0}\;,
\]

\noindent so that ${\cal A}_k(x)$ are three-dimensional
components of the classical Maxwell field (\ref{e3}).
The last formulas establish a correspondence between the
wave functions $f^{k\perp}(x)$ of the transversal photons in the
quantum mechanics and states $|f>$ of the photons in quantum electrodynamics.
One can verify, similar to the massive case, that the actions of the Poincare
generators on the fields (\ref{e4}) and (\ref{e5}) coincide in the both
theories.


\begin{thebibliography}{99}
\bibitem{Landau}L. D. Landau and E. M. Lifshitz, {\em Field Theory}
(Nauka, Moscow, 1973)
\bibitem{BM}F.A. Berezin and M.S. Marinov, JETP Lett. {\bf 21} (1975) 320;
Ann. Phys. {\bf 104} (1977)~336
\bibitem{Casa}R. Casalbuoni, Nuovo Cim. {\bf A33} (1976) 115
\bibitem{BCL}A. Barducci, R. Casalbuoni and L. Lusanna,
Nuovo Cim. {\bf A35} (1976) 377
\bibitem{BDZVH}L. Brink, S. Deser, B. Zumino, P. di Vechia and
P. Howe, Phys. Lett. {\bf B64} (1976) 435
\bibitem{BVH}L. Brink, P. di Vechia and P. Howe, Nucl. Phys. {\bf B118}
(1977) 76
\bibitem{BSSW}A.P. Balachandran, P.Salomonson, B. Skagerstam and
J. Winnberg, Phys. Rev. {\bf D15} (1977) 2308
\bibitem{Sri}P.P. Srivastava, Nuovo Cim. Lett. {\bf 19} (1977) 239
\bibitem{GT}V.D. Gershun and V.I. Tkach, Pis'ma Zh. Eksp. Teor.
Fiz. {\bf 29} (1979) 320
\bibitem{BL}A. Barducci and L. Lusanna, Nuovo Cim.
{\bf A77} (1983) 39; J. Phys. {\bf A16} (1983) 1993
\bibitem{GNR}J. Gomis, M. Novell and K. Rafanelli, Phys. Rev. {\bf D34}
(1986) 1072
\bibitem{HA}P.S. Howe, S. Penati, M. Pernice and P. Townsend,  Phys. Lett.
{\bf B215} (1988) 555; Class. Quant. Grav. {\bf 6} (1989) 1125
\bibitem{Dirac}P.A.M. Dirac, {\em Lectures on Quantum Mechanics}
(Yeschiva University, 1964)
\bibitem{DI}D.M. Gitman and I.V. Tyutin, Class. Quantum Grav.
{\bf 7} (1990) 2131; {\em Quantization of Fields with Constraints}
(Springer, Berlin, 1990)
\bibitem{GG}G.V. Grigoryan and R.P. Grigorian, Vad. Fiz. {\bf 53} (1991) 1062
\bibitem{FW}L.L. Foldy and S.A. Wothuysen, Phys. Rev. {\bf 78} (1950) 29
\bibitem{BS}N.N. Bogoliubov and D.V. Shirkov, {\em An Introduction
to the Theory of Quantized Fields} (John Wiley, New York, 1959)
\end{thebibliography}
\end{document}